\documentclass[aps,prb,twocolumn,groupedaddress,floatfix]{revtex4}

\usepackage{graphicx} % for inserting graphics into figures.

\begin{document}

%\preprint{}

\title{Helium Adsorption in Silica Aerogel near the Liquid-Vapor Critical Point}


\author{Tobias Herman}
\email[]{therman@phys.ualberta.ca}
%\thanks{}
\author{James Day}
%\thanks{}
\author{John Beamish}
\email[]{beamish@phys.ualberta.ca}
%\thanks{}
\affiliation{Department of Physics, University of Alberta,
Edmonton, Alberta, Canada}


\date{\today}

\begin{abstract}

We have investigated the adsorption and desorption of helium near
its liquid-vapor critical point in silica aerogels with porosities
between 95\% and 98\%.  We used a capacitive measurement technique
which allowed us to probe the helium density inside the aerogel
directly, even though the samples were surrounded by bulk helium.
The aerogel's very low thermal conductivity resulted in long
equilibration times so we monitored the pressure and the helium
density, both inside the aerogel and in the surrounding bulk, and
waited at each point until all had stabilized. Our measurements
were made at temperatures far from the critical point, where a
well defined liquid-vapor interface exists, and at temperatures up
to the bulk critical point.  Hysteresis between adsorption and
desorption isotherms persisted to temperatures close to the
liquid-vapor critical point and there was no sign of an
equilibrium liquid-vapor transition once the hysteresis
disappeared.  Many features of our isotherms can be described in
terms of capillary condensation, although this picture becomes
less applicable as the liquid-vapor critical point is approached
and it is unclear how it can be applied to aerogels, whose tenuous
structure includes a wide range of length scales.

\end{abstract}

\pacs{64.60.Fr, 64.70.Fx, 68.03.Cd}
%\keywords{}

\maketitle

\section{Introduction}

Fluid phase transitions can be drastically changed by confinement
in small pores.  Helium has a rich phase diagram and
well-understood bulk behavior and is an ideal system in which to
study these changes.  Silica aerogels, with their tenuous
structure and extraordinarily low densities, provide a unique
opportunity to introduce impurities in a controllable way and so
to address fundamental questions about how disorder, finite size
effects and surfaces affect phase transitions and critical
behavior.  Recent work on the helium/aerogel system has provided
insights into the role of disorder in a wide range of phase
transitions.  For example, the lambda transition for $^4$He in
aerogels~\cite{Chan88,Yoon98} remains sharp, but with a non-bulk
critical exponent for superfluid density, while for $^3$He
aerogels suppress or even completely eliminate
superfluidity~\cite{Matsumoto97,Porto99}, resulting in a zero
temperature ``quantum phase transition.''  For  $^3$He-$^4$He
mixtures, the presence of 2\% or even 0.5\% silica (i.e. aerogels
with porosity of 98\% or 99.5\%) has dramatic
effects~\cite{Chan96,Kim93,Mulders95} on the entire phase diagram,
causing the phase separation curve to detach from the lambda line
and stabilizing a region of dilute $^4$He superfluid inaccessible
in bulk mixtures.

The range of effects aerogels have on the various transitions
reflects differences in their correlation lengths and in how the
order parameter couples to the aerogels. In superfluid $^4$He, the
correlation length is very small and only approaches that of the
aerogel structure very close to the lambda point.  The correlation
length in superfluid $^3$He is much larger, comparable to aerogel
length scales at all temperatures. In $^3$He-$^4$He mixtures, the
aerogel strands couple directly to the order parameter ($^3$He
concentration), in contrast to the much weaker coupling to the
superfluid order parameters in $^3$He and $^4$He.

The liquid-vapor phase transition in porous media has considerable
fundamental and practical interest.   It is often described within
the framework of capillary condensation since, in geometries like
pores or channels, fluids tend to condense more readily than in
bulk.  That is, liquid forms at pressures lower than the bulk
saturated vapor pressure. This behavior is usually described by
taking into account the energetics of liquid-vapor and
liquid-solid interfaces (i.e. surface tension) and the depression
of condensation pressure is fairly well-described by the Kelvin
equation, even in very small pores~\cite{Schmidt95}.  Capillary
condensation is characterized by very deep metastable states
during adsorption and desorption, which leads to hysteresis along
adsorption isotherms.

Although extracting pore size distributions for real porous media
requires simplifying assumptions, the fundamental aspects of
capillary condensation are well explained. However, near the
liquid-vapor critical point (LVCP) this picture breaks down.
Thermal fluctuations grow from atomic to macroscopic scale as the
LVCP is approached and many thermodynamic properties diverge (e.g.
the liquid-vapor interface thickness) or tend to zero (e.g. the
surface tension). Close to the critical point, density
fluctuations might exceed the size of the pores, raising the
question of how the critical behavior responds to confinement. In
bulk fluids, the liquid-vapor transition falls into the Ising
universality class and deGennes~\cite{Brochard83,deGennes84}
suggested that fluids in porous media may provide a realization of
the random field Ising model (RFIM).  However, most measurements
in pores and gels show slow dynamics~\cite{Frisken95b},
metastability and hysteresis and this complicated behavior makes
it difficult to compare directly to theories of phase transitions
in disordered media.

Experimental work on capillary condensation of near-critical
fluids includes dense porous glasses such as CPG (controlled pore
glass)~\cite{Machin99,Thommes94,Thommes95}.  These studies show a
slight narrowing in the phase separation curve, with the ``vapor''
branch shifted to higher densities than in the bulk fluid.  In
addition, the termination of this curve, sometimes referred to as
the ``capillary critical point,'' is shifted to a temperature
below the bulk LVCP.  Despite the relatively well defined pore
geometries of these systems there is no general agreement about
how to picture these systems near the LVCP. Aerogels provide an
opportunity to study this transition in a very different medium,
one without a well-defined pore shape. With their tenuous network
of silica strands, it is hard to imagine a liquid-vapor meniscus
with the uniform negative curvature usually associated with
capillary condensation.  On the other hand, aerogels could be the
ideal system in which to study liquid-vapor critical behavior in
the dilute impurity limit. However, a recent direct measurement of
density fluctuations in carbon dioxide confined in
aerogel~\cite{Melnichenko04} showed little evidence for a
diverging correlation length at the LVCP and thermal fluctuations
may be less important than those introduced by the disorder in the
aerogel structure~\cite{Detcheverry04}.

The first liquid-vapor experiments involving
aerogels~\cite{Wong90-2567} showed a reduced critical temperature
for $^4$He in a 95\% porosity sample and a coexistence region
which was dramatically narrowed (by a factor of ten).  Below the
critical point, isotherms appeared to have discontinuous density
jumps between vapor and liquid-like phases and no hysteresis was
observed.   The surprisingly narrow coexistence curve, determined
from heat capacity and isotherm measurements, could be fit using
the bulk critical exponent. Measurements with N$_2$~\cite{Wong93}
showed slow dynamic behavior (in density fluctuations in light
scattering) and a coexistence curve that was also narrower than
for bulk (although substantially broader than for $^4$He). More
recently, a low frequency mechanical pendulum technique was used
to measure adsorption isotherms and
study~\cite{Gabay00-585,Gabay00-99} the liquid-vapor behavior of
$^4$He in a similar aerogel. The measurements were affected by
very long thermal response times but showed finite isotherm slopes
and hysteresis between filling and emptying, even near the
critical point.  The behavior was reminiscent of capillary
condensation, rather than equilibrium coexistence and critical
behavior.  Our preliminary ultrasonic, acoustic resonator and
isotherm
measurements~\cite{Tan00,Herman02,Herman03,Beamish03-340,Beamish04-339}
on neon and helium in aerogels also showed hysteresis and our
phase separation curves were narrower than in bulk fluids, but
substantially broader than the $^4$He curve in
Ref.~\onlinecite{Wong90-2567}.

Recently, computer models of fluid adsorption in aerogel have also
reached a level of realism which allows direct comparison between
experimental and simulated adsorption
isotherms~\cite{Detcheverry03, Detcheverry04}.  These simulations
were made using a lattice-gas model, with the aerogel structure
produced by a diffusion-limited cluster-cluster aggregation
algorithm.  The use of the lattice-gas model allows the very large
simulation volumes necessary for investigating low density
aerogels. The simulations favor an interpretation without a
macroscopic equilibrium phase transition near the bulk critical
point and provide a picture of the liquid-vapor interface during
adsorption in aerogel.  The simulations show an interface with a
range of curvatures, which becomes less distinct as the
temperature is raised.

The liquid-vapor phase transition has not been systematically
experimentally studied around the LVCP.  While adsorption
isotherms of low density aerogels at temperatures far below the
LVCP have shown how the isotherm shape changes with
porosity~\cite{Tulimieri99}, data near the LVCP are restricted to
a single porosity (namely 95\%). How hysteresis in the adsorption
isotherms disappears as the temperature is raised has not been
quantitatively studied for fluids in aerogel, and raises
interesting questions about the effects of impurities on
liquid-vapor critical behavior.  The existence of equilibrium
liquid-vapor transition for fluids in low density aerogels,
necessary for there to be true critical behavior, is an open
question.  While there is little evidence for liquid-vapor
critical behavior in denser media, as the concentration of
impurities (i.e.\ aerogel density) is reduced this critical
behavior may resurface.

In this paper, we describe a series of experiments on the
isothermal adsorption and desorption of helium in silica aerogels
of different densities.  The aerogels were surrounded by bulk
helium, but we used a capacitive measurement technique which
allowed us to probe the helium density inside the aerogel
directly.  We controlled and varied the pressure, and thus the
chemical potential, rather than admitting fixed amounts of helium.
Thermal equilibration within aerogels was exceptionally slow;
their tenuous microstructure suppressed convection and conducted
heat very poorly.  This slow equilibration was controlled by the
heat of adsorption when fluid is added to (or removed from) the
gel. Pressures equilibrated rapidly due to the aerogels' large
permeability, but the heat of adsorption produced a temperature
gradient which kept the fluid density within the gel from
equilibrating until this heat had been conducted through the
sample. We minimized equilibration times by cutting our samples as
thinly as possible (less than 0.5mm), but the system could still
take hours to equilibrate after a pressure step. Since there was
no way to avoid this slow equilibration in monolithic aerogels, we
monitored the pressure, bulk helium density and helium density
inside the aerogel and waited at each point until all had
stabilized. We then made detailed measurements to see how the
isotherm shapes and hysteresis depended on aerogel density and how
they evolved with temperature near the LVCP.

Our results show that hysteresis in adsorption isotherms persists
to temperatures very close to the LVCP, and there is no sign of an
equilibrium liquid-vapor transition once the hysteresis
disappears. With aerogel samples with porosities between 95\% and
98\%, and isotherms taken at temperatures from 4.880K to helium's
bulk critical point at 5.195K, we are able to analyze the
evolution of adsorption isotherms with temperature and aerogel
density. The shapes of our isotherms agree with simulation
studies~\cite{Detcheverry04}, and recent
experimental~\cite{Gabay00-585} work, but show no signs of the
surprisingly narrow coexistence region~\cite{Wong90-2567} seen in
the first study published on helium condensation on aerogel.

\section{Experiment Details}

\subsection{Aerogel samples}
Adsorption isotherms were collected for helium condensation in
several aerogel samples.  The two samples studied in detail had
densities of $110 \frac{kg}{m^3}$ and $51 \frac{kg}{m^3}$
corresponding to porosities of 95\% and slightly less than 98\%
respectively; throughout this paper they are referred to as
aerogels ``B110'' and ``B51.''  These aerogel samples were
synthesized in our lab from TMOS  using the standard one-step base
catalyzed method followed by supercritical extraction of the
methanol solvent\cite{Poelz82}. Sample B51 was the lowest density
sample that we could reliably get to gel using the one-step
process --- to go to lower densities (and thus higher porosities)
one must use a two-step synthesis procedure\cite{Tillotson92}.
Data from a third sample, designated ``U90,'' are not presented in
detail here since the aerogel was not made in our laboratory and
its synthesis conditions are unknown. Its physical appearance
resembled that of our other two samples, and it had a density of
$90 \pm 25 \frac{kg}{m^3}$.

The aerogel samples were $\sim0.5$mm thick discs
cut\footnote{Aerogel discs were cut with a high speed grinder
(Micro Motor handpiece Model \#MH-135, from Foredom Electric) fit
with a diamond coated disc (Horico SuperDiaflex H635F220, supplied
by Hopf, Ringleb \& Co. GmbH\&Cie.)} from monolithic aerogel
cylinders about 12mm in diameter and $\sim2.5$cm long. Because of
the highly fragile nature of aerogels, great care had to be taken
at all points to minimize any forces on the discs. Copper films
were then deposited directly onto both faces of the aerogel discs
by thermal evaporation through a 9mm diameter mask (deposited at
0.3--0.4 nm/sec to a thickness of $ 200 \pm 20$nm, as determined
by quartz crystal thickness monitor). Empirically, metal did not
seem to penetrate significantly into the aerogel and the copper
film did not plug the pores as it would for a denser porous
medium.  The metal film acted as a continuous electrode, confirmed
by electrical conductivity measurements.  By patterning the
electrodes directly onto to aerogel we ensured that any bulk fluid
in the experimental cell could only influence our measurements
through the capacitor's fringe fields; it could not intrude
directly between the capacitor plates.

The samples acted as parallel plate capacitors, but included
significant effects from fringe fields (electric field lines which
project from the edges of the capacitor) through the bulk helium
in the cell. Since the measured capacitance depended on the
details of all the electric field lines, it was affected by the
environment outside as well as that inside the aerogel. This
effect was clearly seen in the experiment -- for a slice of
aerogel of our dimensions, fringe fields caused the capacitance to
deviate from the ideal infinite parallel plate capacitance by
about 10\%. The aerogel capacitors were also modelled using an
electrostatic modelling program\footnote{``Maxwell$^\copyright$
2D,'' Ansoft Corporation}. The model calculations reproduced the
magnitude of the fringe field effect seen
experimentally~\cite{tobyPhD}.

\subsection{Cell Design}
There were three distinct modules to the cell, each with a single
measurement function -- a central piece and two endcaps. The
center included a capacitive pressure gauge and an inlet capillary
through which fluid could be admitted.  One endcap measured bulk
fluid density using a coaxial cylindrical capacitor with the bulk
fluid acting as a dielectric between the plates. The other endcap
was used to hold the aerogel sample for measurements of fluid
density within the gel.  Details of the experimental setup can be
found in Ref.~\onlinecite{tobyPhD}.

The aerogel lay in a shallow copper cup which made contact with
the bottom electrode on the aerogel slice and was electrically
connected to an isolated feed-through soldered into the cell.
Contact to the top electrode was made by a patch of fine metal
mesh --- the patch acted as a very weak spring, holding the slice
in place while ensuring good electrical contact to another
feed-through.  A typical aerogel sample had a capacitance of $C_0
\approx2$pF. The experimental cell was designed to reduce stress
on the sample, allowing us to work with weak, high porosity,
aerogels.

The bulk density capacitor had a capacitance of
$C_{bulk}\approx35$pF. The outer plate was a 1cm length of 1.25cm
diameter thin-walled stainless steel tubing; the inner plate was a
1cm long solid copper cylinder machined to fit inside the tubing
with a slight gap between the two.  The two plates were separated
by eight small dabs of epoxy --- the epoxy held the plates apart,
but filled a negligible amount of the inter-plate volume (less
than 2\%).

The \textit{in situ} pressure gauge was necessary to obtain
sufficient precision and reproducibility for our experiments. The
gauge was constructed after the design of Straty and
Adams\cite{Straty69}, and consisted of a 0.2mm thick hardened
beryllium-copper diaphragm whose deflection influences the
separation of two parallel brass plates.  The gauge was calibrated
for each data run using a room temperature pressure gauge
\footnote{Mensor 1000psia gauge, Model 4040}.

The cell was mounted on a liquid helium cryostat, with cooling
power provided by helium exchange gas. Temperature was controlled
using a calibrated Germanium resistive thermometer and $200
\Omega$ thick film resistive heater mounted directly on the cell
body. Cell temperature was controlled to $\pm 50 \mu$K with long
term drift also of about  $\pm 50 \mu$K.

\subsection{Data Collection}
Most capacitance data from aerogel B110 were obtained using a
manually balanced capacitance bridge. Since we did not use a low
temperature reference capacitor, the bridge's resolution and drift
limited the experimental resolution to about 1 part in $10^5$.
Later data, including all data on aerogel B51, were collected
using an automated capacitance bridge\footnote{The manual bridge
was a General Radio model 1615-A. The automated bridge was an
Andeen-Hagerling model AH2550A. Both bridges were operated at a
frequency of 1kHz.} with much higher resolution.

Helium was admitted to the cell using two different methods. In
the first method, helium was directly admitted to the cell from a
room temperature gas handling system through a mass flow
controller. This ensured a continuous flow into the cell at a well
controlled rate, but even the slowest rate of flow was too fast to
allow for relaxation within the aerogel. To allow for thermal
equilibration, the flow controller had to be shut off and the
system allowed to relax until the fluid density in the aerogel
stopped changing. Most data from sample B110 were taken in this
manner.

\begin{figure}
\includegraphics{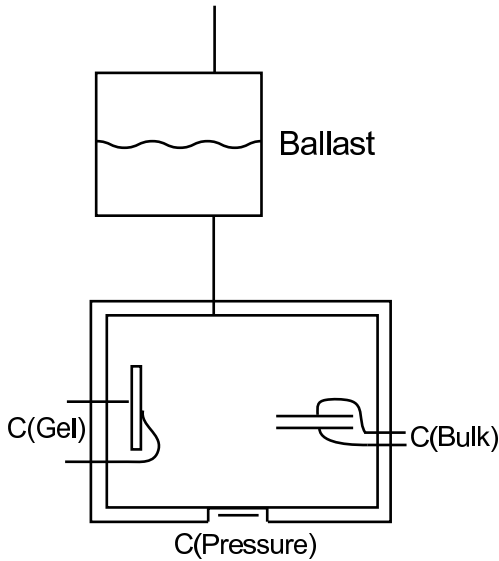}
\caption {\label{IsothermSetupSchematic}A schematic of the low
temperature portion of the setup used to measure aerogel
capacitance during adsorption of helium in silica aerogel.}
\end{figure}

For collecting data on aerogels B51 and U90, a low temperature
ballast volume was mounted about 33cm above the cell for
collecting equilibrated points; the setup is shown schematically
in Fig.~\ref{IsothermSetupSchematic}.
Helium was admitted to the ballast and the pressure of the system
was then controlled by adjusting the temperature of the ballast
volume.  The ballast and cell were connected by a $\sim50$cm
length of 0.2mm i.d. CuNi capillary and their temperatures were
controlled independently. When the ballast held coexisting liquid
and vapor, a small step in the ballast temperature would change
the equilibrium vapor pressure of the helium in the ballast, and
consequently change the pressure in the sample cell.

This latter method is directly analogous to simulation studies of
hysteretic adsorption of fluids in which the chemical potential is
gradually incremented and the response of the fluid density is
monitored.  Using this pressure stepping technique, we ensured
that we remained along the outside edge of any hysteresis loops.
On the other hand, if volumetric bursts were added to, or removed
from, the system, an unknown degree of local heating or cooling
could have occurred within the sample. Thus, precise knowledge
would be lost about where the system was within the hysteretic
region.  When a system is not governed by metastability the
methods should give identical results, but when hysteresis is
present the current state of the system depends on all the states
it occupied along the way.

Using the pressure regulation ballast allowed us to automate small
pressure steps along the adsorption isotherm and required no
additional helium input from the gas handling system outside the
cryostat.  Our technique worked well for isotherms with finite
slopes, but it only produced a few points along very flat
isotherms (i.e.\ isotherms where the aerogel sample filled with
liquid over a narrow pressure range) whereas the dosing technique
would have allowed a higher density of points. For B110 there was
no problem obtaining enough points along the hysteresis loop of
the isotherm, but for B51 the isotherms were so flat that they
would fill or empty completely over a pressure range spanned by
only two or three data points.

The resolution and drift of the manual capacitance bridge were the
limiting factors in the earlier isotherms, but the resolution of
the automated capacitance bridge is about an order of magnitude
better than the experimental noise.  For most of the data
therefore, resolution was set by temperature control. The
resolution of the bulk helium density capacitor is better than
$\pm 0.01 \frac{kg}{m^3}$. The aerogel helium density resolution
is somewhat sample dependent, but always better than $\pm 1
\frac{kg}{m^3}$. Both the bulk and aerogel helium density
capacitances show slight shifts with temperature which contribute
an additional error of less than $0.5 \frac{kg}{m^3}$ over the
temperature range of these data. The method we used for converting
aerogel capacitance to helium density also introduced errors on
the order of those due to electrical noise.

Since the sample was mounted with its axis horizontal, there was a
gravitational  pressure head of up to 60$\mu$bar across the 1.2cm
diameter sample. The uncertainty and drift in cell and ballast
temperatures ($\pm 50 \mu K$), and the slope of the coexistence
curve of helium near its critical point ($\sim 1.5bar/K$) combined
to restrict our experimental pressure resolution to about
70$\mu$bar. The similar size of these two factors meant that both
the sample dimensions and temperature control would have to be
changed to improve experimental resolution significantly.

\section{Results}

\subsection{Bulk Helium Density} The bulk helium capacitor was
calibrated by using the measured capacitance and literature
values\cite{NISTWebBook} for the density of helium liquid and
vapor at saturated vapor pressure at 4.400K. These values were
used to fit the helium density ($\rho_{bulk}^{He}$) to an
expression of the form: \begin{equation} \rho_{bulk}^{He} = A (C -
C_{0}^{eff} )
\end{equation}
where $C$ was the measured capacitance and $A$ and $C_{0}^{eff}$
were adjustable constants which were recalculated for each data
run. The bulk helium capacitor had a slight background temperature
dependence so a small empirical temperature correction was added
to minimize deviations from literature values at higher
temperatures.  A comparison between densities calculated from our
experimental data and values extracted from the NIST Chemistry
WebBook\cite{NISTWebBook} showed deviations less than $ 0.3
\frac{kg}{m^3} $ for temperatures from 4.400K to just below the
bulk critical temperature of helium (5.195K). Since bulk helium
density is only used to account for the effect of the aerogel
capacitor fringe fields, this degree of accuracy is sufficient.

\begin{figure}
\includegraphics[width=\linewidth]{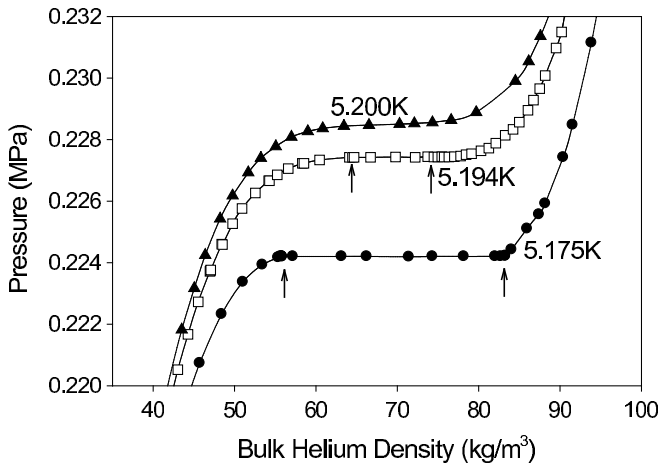}%
\caption {\label{BulkHeliumIsotherms}Isotherms for bulk helium
near the critical temperature of 5.1952K. Below T$_c$ there exists
a region of two-phase coexistence where the isotherm is completely
flat, indicated by the arrows along the isotherms at 5.175K and
5.194K. Above T$_c$ this flat region is replaced by an inflection
point.}
\end{figure}

Using this calibration, bulk pressure-density isotherms were
collected.  Three  examples of bulk isotherms are shown in
Fig.~\ref{BulkHeliumIsotherms}, two below T$_c$ (which was
empirically determined to be 5.1952K on our thermometer
calibration) and one above it. For the T$<$T$_c$ isotherms a
distinct coexistence region is marked by arrows. The width of the
two-phase coexistence region shrinks as the critical temperature
is approached; above T$_c$ there is no coexistence and isotherms
are single smooth curves with an inflection point at the critical
density.  Even when the onset of coexistence was a little unclear
on a pressure-density plot, such as the T=5.194K isotherm in
Fig.~\ref{BulkHeliumIsotherms}, it was obvious when looking at the
time dependence of the data during a constant flow adsorption
isotherm --- once coexistence was reached the capacitance of the
bulk helium capacitor remained constant for a long time, until the
liquid-vapor meniscus in the cell reached the bulk helium
capacitor. From the slopes of supercritical isotherms it is also
possible to determine the compressibility of bulk helium, with
flat isotherms below the LVCP corresponding to a diverging
compressibility.

\subsection{Helium Density in Aerogel}
The effective dielectric constant of a porous medium depends on
the dielectric constant of the matrix, that of the fluid in its
pores, and the geometry of the system.  With knowledge of those
three factors it is possible to calculate an effective dielectric
constant\cite{Pelster99}. Since aerogel acts as a dilute impurity,
it turns out that there is little deviation from a linear
relationship between fluid density and capacitance.  When a single
fluid phase is present there is a nonlinear term in the dielectric
constant of the aerogel which depended on the aerogel porosity
($\phi$) and the fluid dielectric constant ($\epsilon_{He}$). This
term, proportional to $(1-\phi)*(\epsilon_{He} -1)$, was
negligible in our measurements since we were using 95\% and 98\%
porous gels and the dielectric constant of helium is less than
1.05 at the temperatures and pressures used in our experiments.
When phase separated liquid and vapor were present inside the
aerogel, more assumptions had to be made since the geometry of the
liquid-vapor interface was not known. However, deviations from
linearity were still smaller than 3\%\cite{tobyPhD}, less than
both the experimental variations between runs and the features we
were investigating.

There are actually three contributions to the measured capacitance
of the aerogel sample: the capacitance of the empty aerogel
($C_0$), the capacitance added by the helium in the aerogel
($C_{gel}^{He}$), and the capacitance added by the bulk helium in
the fringe fields outside of the aerogel sample ($C_{bulk}^{He}$):
\begin{equation} C_{total} = C_0 + C_{gel}^{He} + C_{bulk}^{He}
\end{equation}

\begin{figure}
\includegraphics[width=\linewidth]{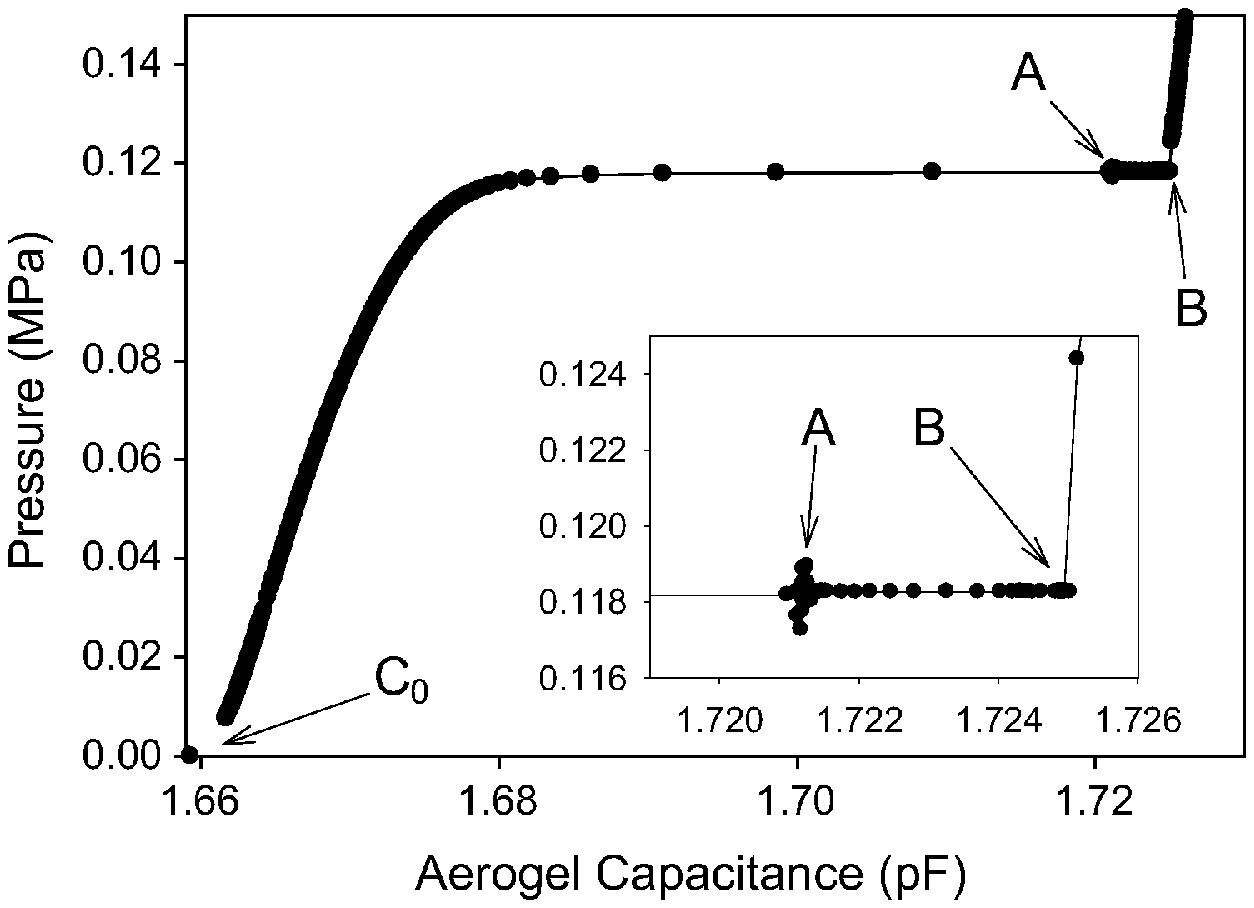}
\caption{\label{95GelFringeEffects}Continuous flow helium
adsorption isotherm for sample B110 at 4.400K. C$_0$ is the
capacitance of the empty aerogel, point ``A'' corresponds to the
aerogel being full of liquid coexisting with bulk vapor outside
the gel, and point ``B'' corresponds to a full gel surrounded by
bulk liquid helium. The inset enlarges the region where fringe
field effects are most obvious.}
\end{figure}

To find the density of helium in the aerogel these three factors
were determined separately, using an adsorption isotherm taken at
4.400K  to determine their relative sizes. The capacitance of the
aerogel was measured when empty (C$_0$), when full with liquid but
surrounded by bulk vapor (C$_A$), and when full with liquid and
surrounded by bulk liquid (C$_B$) --- see
Fig.~\ref{95GelFringeEffects} for details. As the aerogel fills
with helium, capacitance increases until point ``A,'' where the
aerogel is completely filled with liquid.  The spike of noise in
pressure at point A is related to momentary temperature control
problems in the cell as bulk liquid begins to condense into the
cell. As the bulk liquid meniscus moves up over the aerogel sample
(i.e. between points ``A'' and ``B'') the capacitance increases
due to fringe fields in the bulk helium, not due to the helium
density within the sample.  By quantifying these fringe effects it
is possible to correct the capacitance data for the small spurious
shifts due to changes in bulk fluid density in the cell.

We assume that far from the LVCP the density of liquid helium in
aerogel is equal to that of bulk liquid helium; while there may be
a slight enhancement of helium density due to denser layers
adsorbed on the silica strands, it will not be important far from
the critical point.  The first layer of helium  adsorbed on the
aerogel surface accounts for less than 8\% of the open volume in
the aerogel (for 95\% porosity aerogel).  Even if we assume that
the first layer is 25\% denser than bulk liquid helium, then the
total helium density of the system will only be enhanced by about
2\% far from the LVCP.

Using the three points along the 4.400K isotherm mentioned above,
and assuming that helium density at 4.400K in aerogel is identical
to bulk helium density, it is possible to fit the density of the
helium to an expression of the form: \[ \rho_{gel}^{He} = D \left(
C_{gel} - C_0 - \frac{\rho_{bulk}}{E} \right) \] where ``D''
reflects the sensitivity of the capacitor to the density of helium
in the aerogel and ``E'' reflects the effects of the fringe fields
(i.e. the influence of the environment surrounding the gel, which
must be subtracted from the signal). The constants (C$_0$, D, and
E) must be determined for each run since even a small change in
position of the sample can have significant effects on the density
calibration. When bulk capacitance and aerogel capacitance were
measured simultaneously it was simple to account for the fringe
effects.  However, occasionally bulk data were not available, so
the density for bulk helium at SVP was used to approximate
$\rho_{bulk}$ (bulk vapor density, since our isotherms were taken
below bulk saturation).

\subsection{Thermal Equilibration}
Aerogels are truly phenomenal thermal insulators, a fact which
leads to very slow thermal equilibration within samples. By using
the thinnest samples (only a few hundred micrometers thick) we
minimized the thermal equilibration time. To ensure that we were
measuring rate--independent isotherms, we directly monitored the
equilibration of helium density during adsorption and desorption.

\begin{figure}
\includegraphics[width=\linewidth]{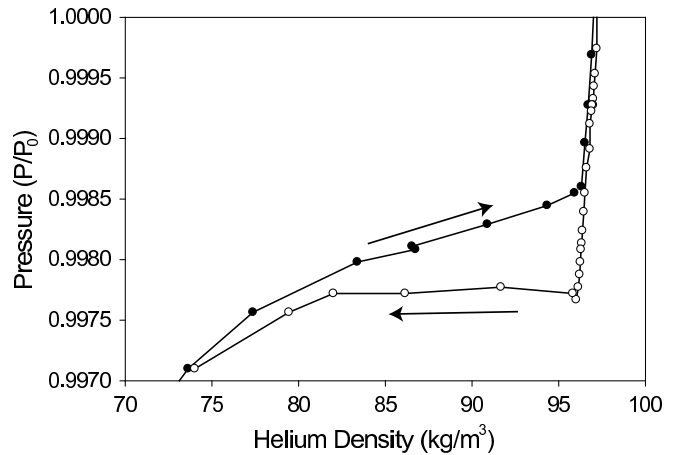}
\caption{\label{B110510K}Equilibrated isotherm at 5.100K in
aerogel B110. This isotherm displays a well defined hysteresis
loop --- thermal equilibrium of points along the hysteresis loop
was very slow, while points that did not lie along the hysteresis
loop equilibrated in a few minutes.  Points along the adsorption
isotherm are shown as solid symbols, while points along the
desorption isotherm are shown as open symbols.}
\end{figure}

\begin{figure}
\includegraphics[width=\linewidth]{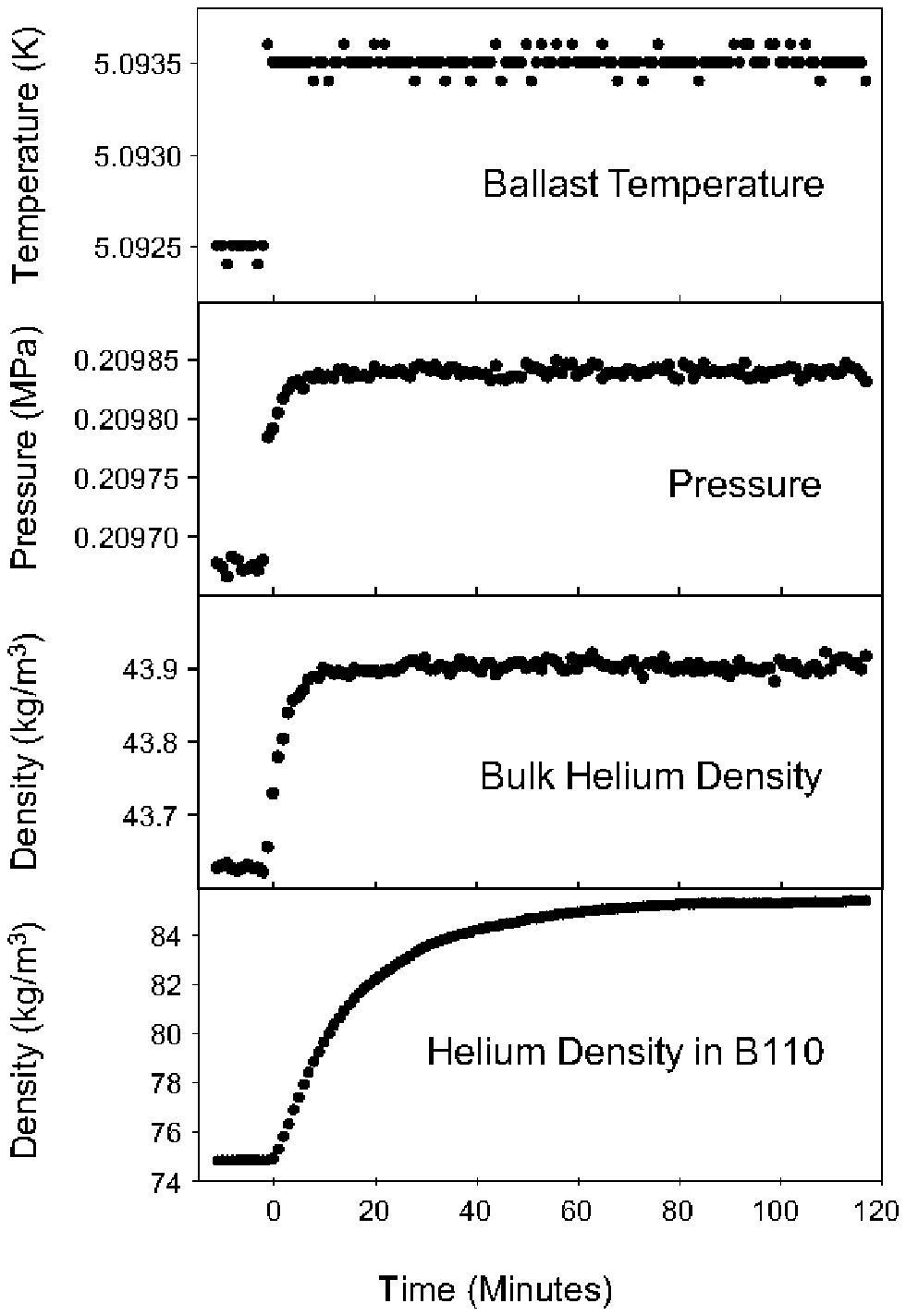}
\caption{\label{B110Equil}Equilibration at 5.100K following a 1mK
step in ballast temperature.}
\end{figure}

\begin{figure}
\includegraphics[width=\linewidth]{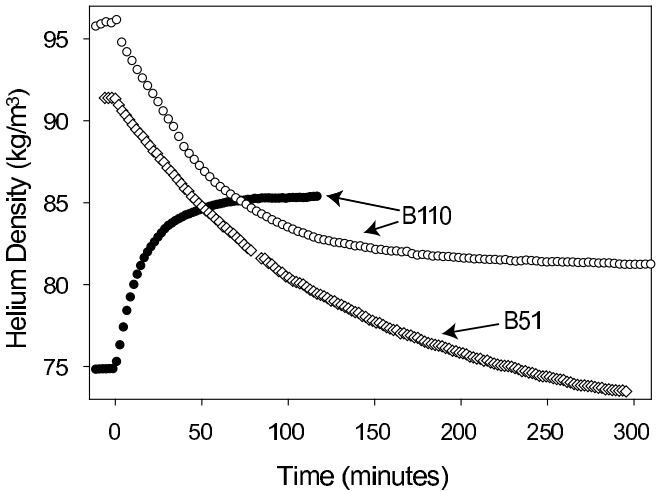}
\caption{\label{95to98Gel51KRelaxEmptying}Comparison between
relaxation in aerogels B110 (filling and emptying) and B51
(emptying) at T=5.100K. The data shown here corresponds to ballast
temperature steps of $\pm1$mK for B110 and $-0.1$mK for B51.}
\end{figure}

One example of an equilibrated isotherm is shown in
Fig.~\ref{B110510K}
for helium in aerogel B110 at 5.100K.
Figure~\ref{B110Equil}
shows the response at 5.100K to a change of 1mK in ballast
temperature. The data in Fig.~\ref{B110Equil} were taken during a
data run subsequent to that in Fig.~\ref{B110510K}. While the step
in ballast temperature was effectively instantaneous and the cell
pressure and bulk helium density responded completely within a few
minutes, the density of helium in the aerogel sample took hours to
equilibrate.  This is probably due to the slow thermal equilibrium
within the aerogel; while the aerogel is highly permeable to
helium, convection is suppressed by the aerogel strands.
Equilibration during emptying took about twice as long as filling,
as shown in Fig.~\ref{95to98Gel51KRelaxEmptying}.

The relaxation in aerogel B51 was even slower than in aerogel
B110; a direct comparison of emptying steps for the two is also
shown in Fig.~\ref{95to98Gel51KRelaxEmptying}. The step along the
B51 desorption isotherm corresponded to a change in T$_{Ballast}$
of only 0.1mK, as opposed to the 1.0mK change shown for aerogel
B110. Since the change in helium density in aerogel B51 is driven
by a much smaller temperature change in the ballast, the pressure
and temperature gradients across the sample during desorption are
smaller in aerogel B51. For a given density change the amount of
latent heat which must be supplied during desorption is
approximately the same in B51 and B110, so the smaller temperature
gradient present in B51 results in slower relaxation. In practical
terms, this means that the equilibration time is inversely related
to the slope of the adsorption isotherm --- the flatter the
isotherm, the slower the equilibration at each pressure.

\begin{figure}
\includegraphics[width=\linewidth]{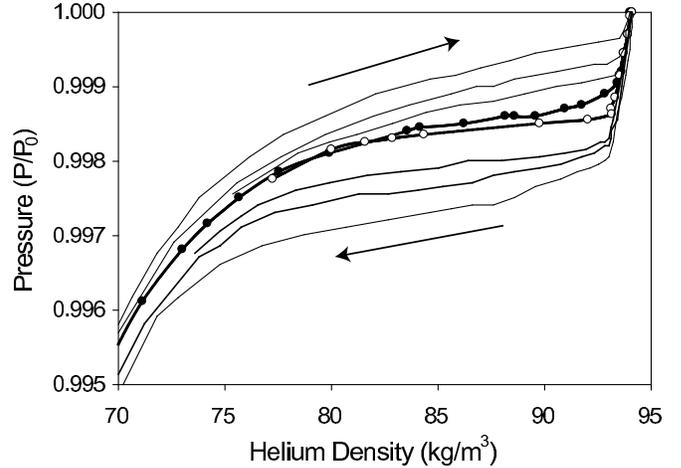}
\caption{\label{95Gel514RateHyst}Isotherms for helium adsorption
in aerogel B110 at 5.140K showing three continuous filling rates
and also equilibrated points. The time to complete the isotherms,
from the outer loop to the inner loop, were 1.5hours, 2.5hours,
and 5hours. The equilibrated points, shown as the innermost loop,
were collected over about 10 hours.}
\end{figure}

Even at the slowest rates possible with our flow controller,
continuous flow adsorption isotherms did not allow for thermal
equilibration along hysteretic isotherms. The effects of filling
rate are illustrated in Fig.~\ref{95Gel514RateHyst}
for isotherms taken at 5.140K in B110. The loop remained when the
density was allowed to equilibrate at each point, but the size and
shape of the continuous flow loops depended sensitively on the
filling and emptying rates. The effect of filling rate was greater
for the lower temperature isotherms while above bulk T$_c$
equilibration was relatively fast and there was no hysteresis
during continuous flow measurements.

\section{Adsorption Isotherms}
\begin{figure}
\includegraphics[width=\linewidth]{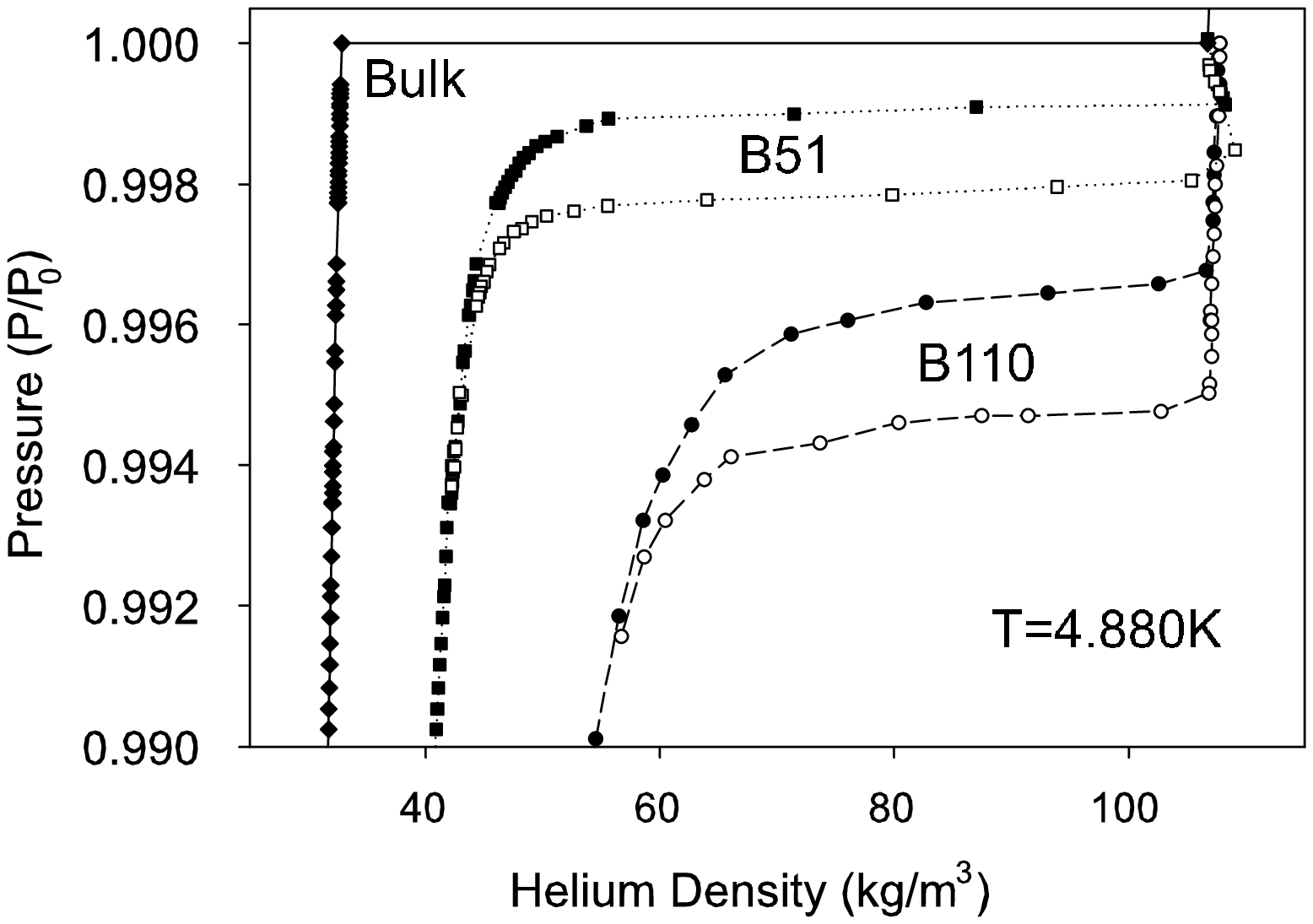}
\caption{\label{95to98Gel488K}Comparison between isotherms in
aerogels B110 and B51 at T=4.880K}
\end{figure}

\begin{figure}
\includegraphics[width=\linewidth]{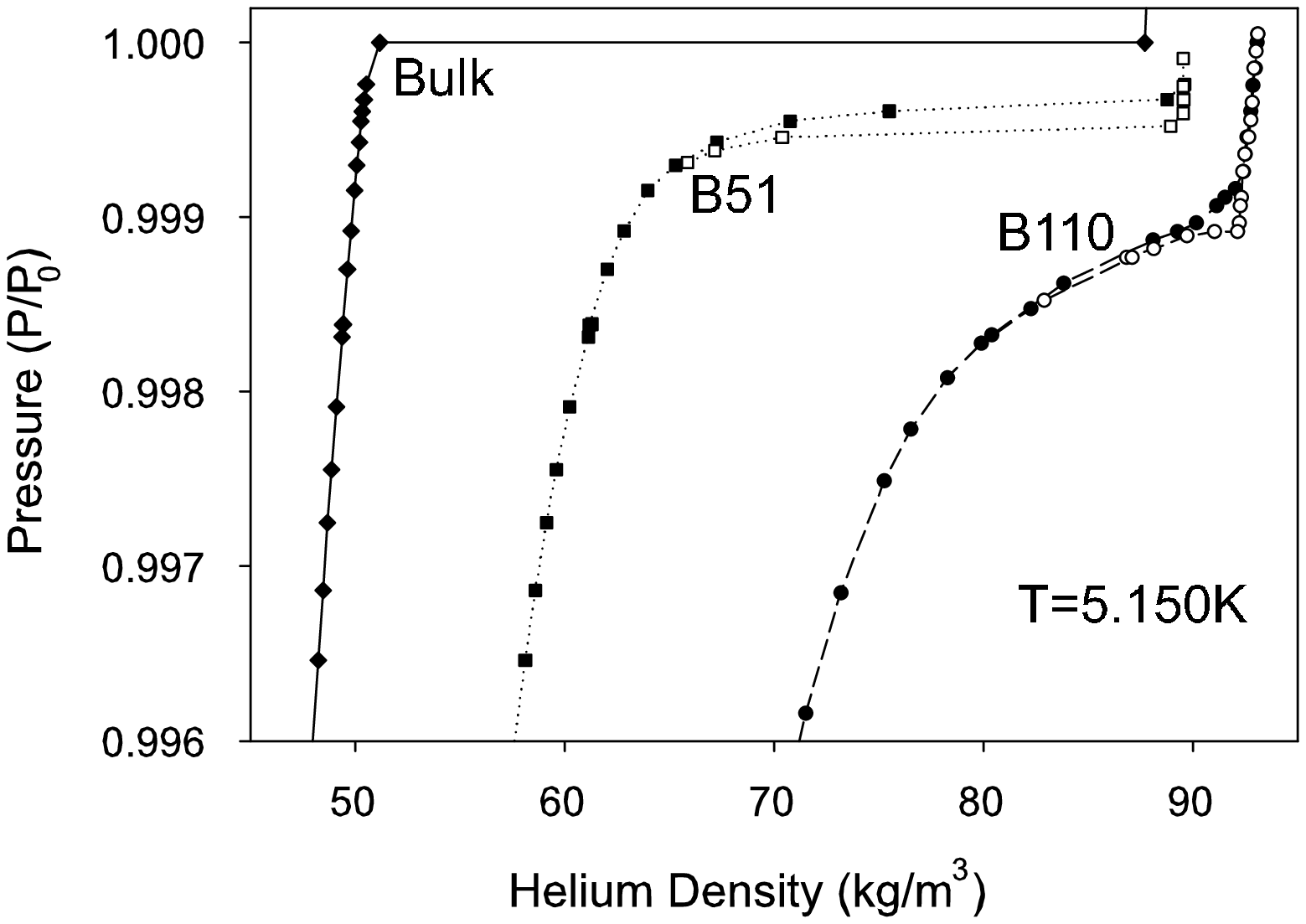}
\caption{\label{95to98Gel515K}Comparison between isotherms in
aerogels B110 and B51 at T=5.150K}
\end{figure}

Adsorption isotherms were collected in aerogels B110 and B51 over
similar temperature ranges; isotherms are shown for T=4.880K and
T=5.150K in Figs.~\ref{95to98Gel488K} and~\ref{95to98Gel515K},
which also include the corresponding measured isotherms for bulk
helium.

Given the similar synthesis conditions, the physical
microstructure of the gels should be very similar, leading to
adsorption isotherms sharing many characteristics. At the
temperatures shown here both samples exhibited hysteresis loops
--- these loops remained resolvable until quite near to the bulk
critical temperature.  The low density sides of the adsorption
isotherms curve smoothly into the hysteresis loops, making it
difficult to precisely identify the onset of capillary
condensation or define ``vapor branches'' for the isotherms.  On
the other hand, the high density sides of the isotherms had sharp
transitions, allowing one to pinpoint the completion of capillary
condensation with more precision.

However, there are also significant differences between adsorption
isotherms in the two aerogels. In aerogel B51 the hysteresis loops
occur closer to saturated vapor pressure (P$_0$) and cover a
larger density range. The low density portion of the adsorption
isotherms in B51 shows less density enhancement before the onset
of capillary condensation --- the less dense medium perturbs the
helium density less from its bulk value. Similarly, the density
enhancement when the gel is full (at 5.150K) is smaller in B51.
Far from the critical point we assumed that the aerogel strands
only perturb the density of the layer of helium directly adjacent
to the silica strands, but by 5.150K the system is close enough to
the LVCP that the liquid-vapor interface has become less distinct
and the fluid is very compressible. In this temperature region,
perturbations in fluid density caused by the silica strands
propagate further into the fluid.

The evolution and disappearance of the hysteresis loop as
temperature is raised is also very different in these two samples.
In B110 the loop disappears between 5.150K and 5.160K ($\sim$40mK
below T$_c$), while in B51 it disappears between 5.170K and 5.180K
($\sim$20mK below T$_c$). Thus, the temperature difference between
the point at which the hysteresis loop closes and the bulk T$_c$
is roughly proportional to the aerogel density.  The manner in
which the loops close also differs. The hysteresis loop in sample
B51 covers a smaller pressure range as the temperature is raised,
but its shape does not change drastically. Sample B110, on the
other hand, exhibits a hysteresis loop that seems to ``zip''
closed as the temperature is raised.

\subsection{Helium in Aerogel B110}

Sample B110 had a porosity close to the aerogel used by Wong and
Chan in their work near the liquid-vapor critical
point\cite{Wong90-2567} and very close to that used by the
Grenoble group\cite{Gabay00-585}. We have equilibrated adsorption
isotherms for sample B110 at temperatures from 4.880K to above the
bulk critical temperature. At temperatures far below the LVCP,
these isotherms show large hysteresis loops, which gradually
shrink as the temperature is raised, until (at about 5.155K) they
finally disappear.  The hysteresis covers a large density range --
much wider than the coexistence curve mapped out by Wong and
Chan\cite{Wong90-2567} close to the LVCP -- but are narrower than
the bulk helium coexistence curve. For all loops, both adsorption
and desorption branches have finite slopes; there are no isotherms
that exhibit the flat coexistence regions seen in bulk helium.

\begin{figure}
\includegraphics[width=\linewidth]{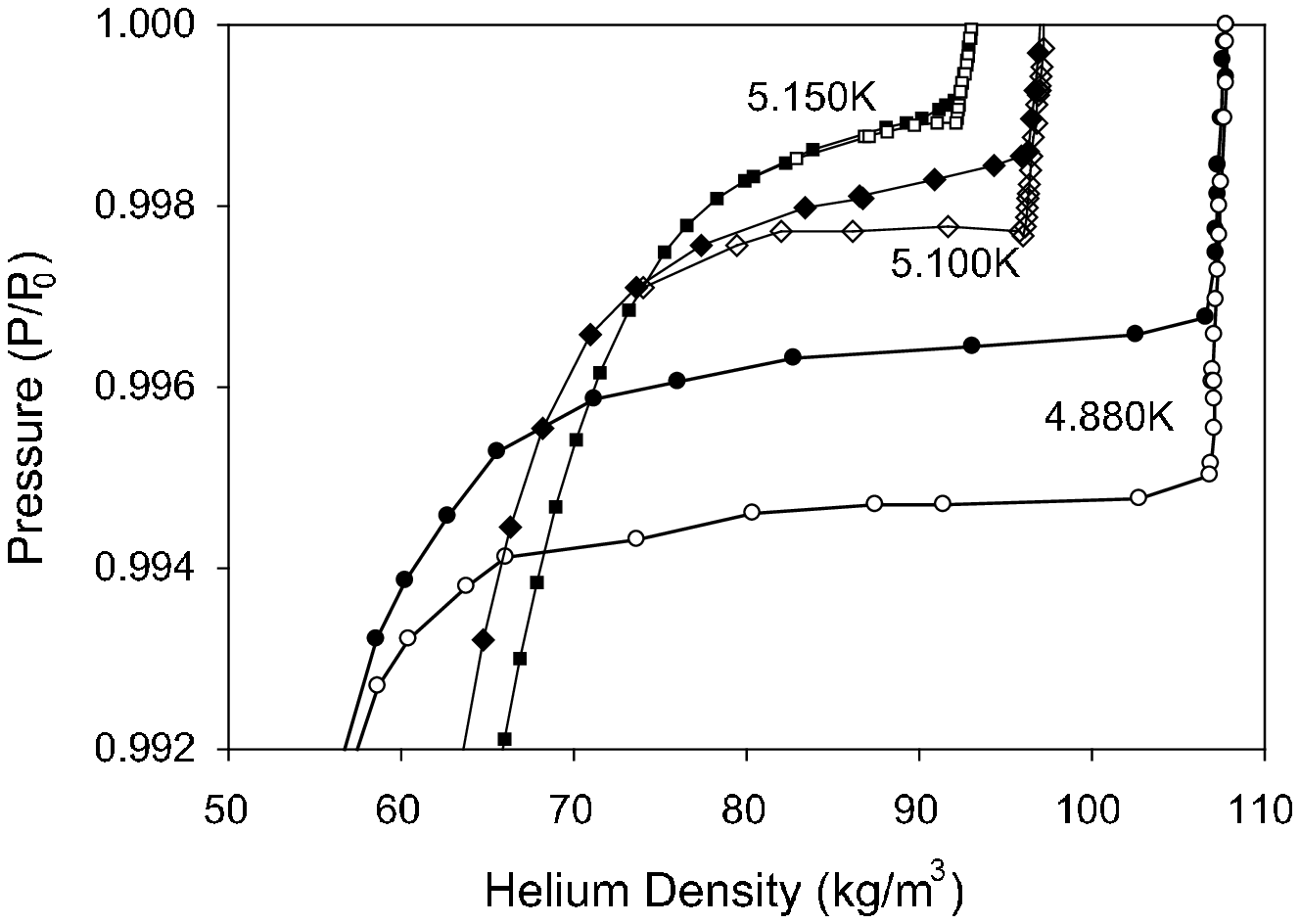}
\caption{\label{95GelLowT3Isotherms}Three isotherms in aerogel
B110 at temperatures: T=4.880K, 5.100K, and 5.150K.}
\end{figure}

\begin{figure}
\includegraphics[width=\linewidth]{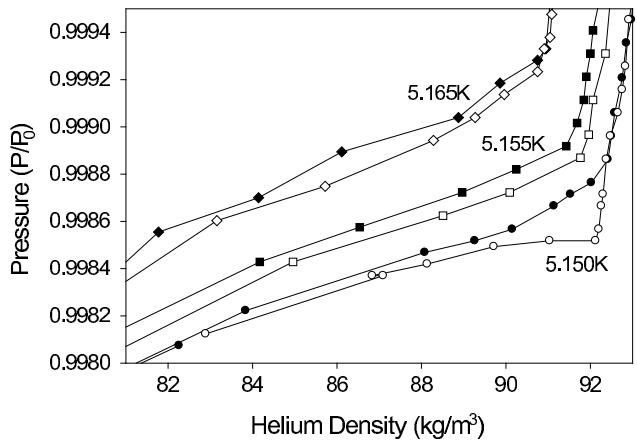}
\caption{\label{95GelMidT3Isotherms}Isotherms in aerogel B110:
T=5.150K, 5.155K, and 5.165K.  The data have been shifted to make
this figure clearer --- the T=5.155K pressure data have been
shifted by $-0.0002$~P/P$_0$ while the T=5.150K pressure data have
been shifted by $-0.0004\frac{P}{P_0}$.}
\end{figure}

The evolution with temperature of the hysteresis loops in B110 is
shown in Fig.~\ref{95GelLowT3Isotherms}
--- they shrink and become more triangular as the temperature is
raised.  The high density end of the loop remains well defined,
especially along the desorption branch, but the hysteresis loop
becomes less and less distinct at the low density end. The loop
transforms from a roughly rectangular shape to a triangle until,
at higher temperatures, it completely disappears
(Fig.~\ref{95GelMidT3Isotherms}).  While the hysteresis loop is
still distinct at 5.150K, by 5.155K any loop is too small to
resolve.  The slight difference between the adsorption and
desorption isotherms at 5.155K and 5.165K in this figure is a
measure of the resolution and drift in our system --- there does
not appear to be any distinct loop as there was for the lower
temperatures.

Since the mechanism for hysteresis in aerogel is not known, the
disappearance of a hysteresis loop above 5.150K need not imply the
disappearance of distinct pore liquid and pore vapor phases.  A
fairly sharp kink at the high density side of the isotherm remains
after hysteresis has disappeared. The eventual disappearance of
this kink at some temperature above the disappearance of
hysteresis but below the bulk critical temperature may be an
indication of the disappearance of a distinct ``pore liquid''
phase.

\subsection{Helium in Aerogel U90}
\begin{figure}
\includegraphics[width=\linewidth]{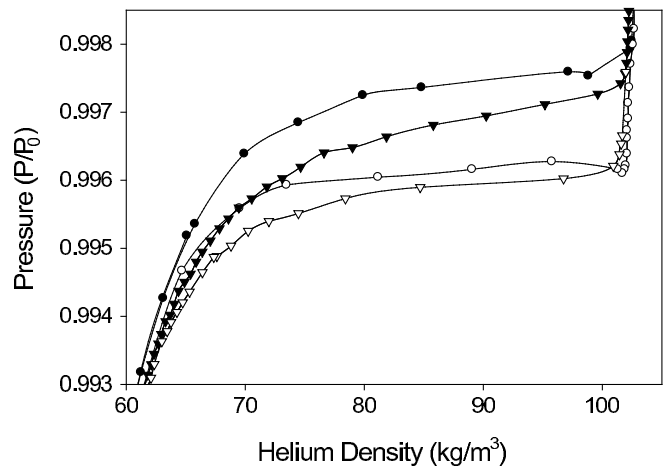}
\caption{\label{B110toU1005K}Isotherms at 5.000K in aerogels
B110(circles) and U90(triangles).  The B110 data were collected
manually while U90 data were collected using the automated system
incorporating the pressure regulation ballast.}
\end{figure}

We have also collected a series of isotherms on another aerogel
sample, designated ``U90,'' with a density of about $90 \pm 25
\frac{kg}{m^3}$.  Data were collected using the improved system,
including the pressure regulation ballast.  A single isotherm from
this sample is included in Fig.~\ref{B110toU1005K}.  The
adsorption and desorption isotherms in aerogel U90 were very
similar to those in B110, including their temperature dependence.

\subsection{Helium in Aerogel B51} 
\begin{figure}
\includegraphics[width=\linewidth]{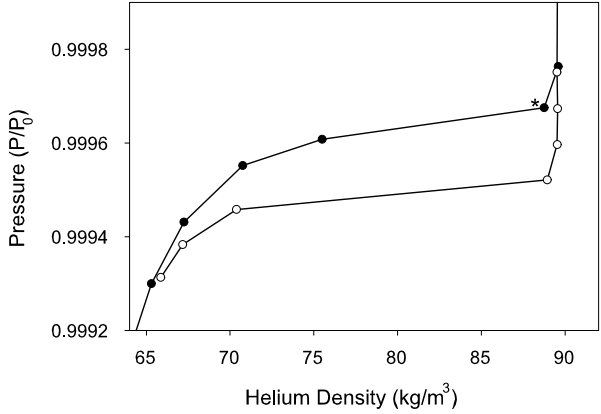}
\caption{\label{98Gel515Isotherm}Adsorption isotherm in aerogel
B51 at T=5.150K. The only point in the plot which does not
represent a fully relaxed state is indicated by the asterisk(*).
Note that this loop is only a few times wider than our pressure
resolution.}
\end{figure}

Sample B51 had a density of $51
\frac{kg}{m^3} $, corresponding to a porosity of slightly less
than 98\%.  Capillary condensation occurred at higher relative
pressures than in the denser gel and the gel filled (or emptied)
over an exceptionally narrow pressure range. Isotherms at 4.880K
and 5.150K were shown earlier (Fig.'s~\ref{95to98Gel488K}
and~\ref{95to98Gel515K}). A single isotherm at 5.150K is shown in
greater detail in Fig.~\ref{98Gel515Isotherm}.  This can be
compared to the 5.100K isotherm for B110 (Fig.~\ref{B110510K}).
Most of the points along the 5.150K isotherm have been allowed to
relax completely, although there was a single point, marked by an
asterisk (*) in Fig.~\ref{98Gel515Isotherm}, that had not quite
equilibrated.

\begin{figure}
\includegraphics[width=\linewidth]{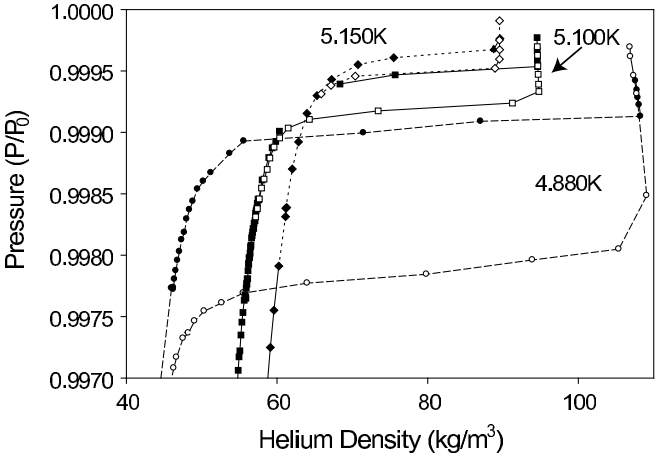}
\caption{\label{98GelLowT3Isotherms}Three isotherms in aerogel
B51: T=4.880K, 5.100K, and 5.150K.  The data along the 4.880K
hysteresis loop is slightly out of thermal equilibrium along the
flattest sections.}
\end{figure}

\begin{figure}
\includegraphics[width=\linewidth]{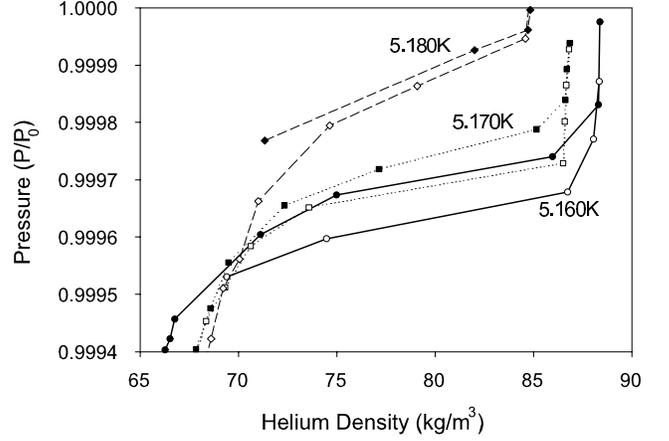}
\caption{\label{98GelHighT3Isotherms}Three isotherms in aerogel
B51: T=5.160K, 5.170K, and 5.180K.  Some points have been
eliminated from the isotherm at 5.180K because they appeared to
drift over time, indicating that we are operating at the limit of
resolution for this system.}
\end{figure}

Figures~\ref{98GelLowT3Isotherms} and \ref{98GelHighT3Isotherms}
show the evolution of the hysteresis loops with temperature.
All isotherms are constructed from equilibrated points except the
4.880K isotherm --- because of the extremely slow thermal
equilibration within sample B51, it was not practical to wait for
the points along the flattest region of the hysteresis loop to
completely equilibrate at this one temperature.  The hysteresis
loops extend over a larger range of densities than in B110 but the
aerogel fills or drains almost entirely in a single $\sim 100 \mu
bar$ pressure step. Hysteresis is still resolvable in the isotherm
at 5.170K, although the precision of our pressure control is
barely good enough (a few parts in $10^5$) to resolve the loop.
At 5.180K no hysteresis was seen.  A number of points along the
higher temperature isotherms showed a slow drift that could
sometimes be comparable to the density change seen during that
isotherm's step. This is an indication that at such temperatures
the isotherms are \emph{very} sensitive to cell temperature and
pressure and we have reached the limits of our ability to control
the system.

\section{Discussion}

Previous measurements\cite{Wong90-2567,Wong93,Gabay00-585} of
liquid-vapor behavior have been made using aerogels with similar
density to our sample B110 ($110 \frac{kg}{m^3}$, 95\% porosity)
and we can compare our results to those experiments. In this
aerogel, our isotherms (Figs.~\ref{95GelLowT3Isotherms}
and~\ref{95GelMidT3Isotherms}) have nearly rectangular hysteresis
loops far from the LVCP, but these become more triangular at
higher temperatures. The loops finally disappear about 40mK below
the bulk LVCP.  All isotherms have non-zero slopes, in contrast to
the flat region associated with liquid-vapor coexistence in bulk
helium (e.g. in Fig.~\ref{95to98Gel515K}). The onset of capillary
condensation along the isotherms is gradual, making it impossible
to unambiguously identify a low density fluid phase in the aerogel
that is analogous to bulk vapor.

Gabay \textit{et al}.~\cite{ClaudeGabayPhd,Gabay00-585} used a
mechanical oscillator to measure the helium density while
isothermally flowing helium at different rates into or out of a
similar aerogel. The hysteresis they observed depended on the flow
rate, but was significant even for filling times as long as 64
hours. Their highest temperature hysteresis loops, at 5.140K,
closely resemble those shown in Fig.~\ref{95Gel514RateHyst}. Our
measurements show that this hysteresis persists in the static
limit, implying that there are metastable thermodynamic states for
helium in aerogel, even very near the bulk critical point.  The
rate dependence they observed is consistent with our long
equilibration times, given their larger sample dimensions (1cm
versus our 0.5mm).  They also found that the isotherms always had
finite slopes, in agreement with our measurements.

Wong and Chan, on the other hand, saw qualitatively different
behavior~\cite{Wong90-2567} near the LVCP of helium.  They
measured heat capacity along isochores and used the temperatures
of the peaks to map out an extremely narrow coexistence curve near
the LVCP, with the liquid branch shifted to \emph{lower} density
than bulk liquid in contrast to our measurements. They
supplemented these measurements with several adsorption isotherms
which had flat regions consistent with coexistence, and reported
no hysteresis. Their isotherm measurements had limited pressure
resolution, so would not have resolved the smallest slopes we
observed.  The absence of hysteresis in these isotherm is
puzzling, but could reflect differences in how our samples were
filled.   They used the more common method in which known
quantities of gas are admitted in bursts and then let come to
equilibrium --- in essence stepping density rather than pressure.
However, this can cause significant local heating as the gas
condenses into the sample. When condensation occurs over a very
narrow pressure range (as it does in aerogels near the critical
point) a local temperature jump of even a few millikelvin would
correspond to relative pressure shifts larger than the width of a
hysteresis loop itself.  The system would then relax along a
different thermodynamic path and might not end up in the same
metastable state it would reach if the chemical potential was
isothermally changed in steps, as in our measurements, or
continuously, as in Gabay \textit{et al}.'s work. At lower
temperatures, all techniques show hysteretic adsorption loops that
cover a relatively wide density range, like the 4.880K data shown
in
Fig.~\ref{95GelLowT3Isotherms}~\cite{ClaudeGabayPhd,Tulimieri99}.

The very narrow coexistence curve that Wong and Chan deduced from
their heat capacity measurements was a surprising result.  Neither
we nor Gabay \textit{et al}.\ were able to find either a true
coexistence region or any features in our isotherms that would
produce such a narrow curve.  In addition to the method used to
admit helium, there are several differences between the
experiments which may be relevant. Both our cell and that of Gabay
\textit{et al}.\ had large open volumes, so the aerogel was always
in contact with a substantial reservoir of bulk helium; Wong and
Chan's cell had much less bulk volume.  Our measurements and those
of Gabay \textit{et al}.\  were done isothermally; Wong and Chan's
heat capacity measurements followed isochores. These differences
mean that less helium was adsorbed or desorbed following a
temperature step in their isochoric measurements than following a
pressure step in our isotherms. However, even with no bulk helium
present, a temperature step produces gradients in temperature and
thus in pressure, resulting in an internal redistribution of
helium within the aerogel, with similar long equilibration times.
Since our sample dimensions were comparable to those in Wong and
Chan's heat capacity measurements (about 0.25mm thick in their
case), their samples probably had similarly long time constants,
up to several hours. This makes interpretation of AC heat capacity
measurements difficult, since only a very thin layer of helium
near the surface of the sample would respond even at a heater
frequency as low as 0.1 Hz.

Although our sample density is similar to that used by Wong and
Chan, the liquid-vapor phase behavior could be sensitive to small
differences in porosity or microstructure.  However, we made
measurements in another aerogel, U90, with similar density ($90
\frac{kg}{m^3}$) but synthesized elsewhere and
Fig.~\ref{B110toU1005K} shows that the behavior at 5.000K is
essentially the same as the behavior of B110 (as was the evolution
of the isotherms with temperature).  Thus, two different gels with
similar densities, but from different sources, showed remarkably
similar behavior.  This supports the position that it is unlikely
that any small structural differences between our samples and that
of Wong and Chan could account for the dramatically different
behavior observed in our experiments.

Our measurements on sample B51 ($51 \frac{kg}{m^3}$, 98\%
porosity) show how the shape and temperature dependence of
adsorption isotherms depend on aerogel porosity, as summarized in
Figs.~\ref{95to98Gel488K} and~\ref{95to98Gel515K}.   As expected,
the effects of the aerogel on helium's liquid-vapor behavior
become smaller as the density of the aerogel is reduced. Isotherms
in the more porous gel (B51) resemble those in B110, but
condensation occurs over a narrower pressure range, closer to bulk
saturation. Hysteresis loops extend over a larger density range
and persist to even higher temperatures, finally disappearing
about 20mK below the bulk critical temperature. Perhaps more
significantly, the shape of the hysteresis loops does not change
as the LVCP is approached (Figs.~\ref{98GelLowT3Isotherms}
and~\ref{98GelHighT3Isotherms}); they remain rectangular, in
contrast to the denser aerogel where they become triangular. The
low density onset of condensation is sharper, but still rounded,
and isotherms have a smaller, but non-zero slope at all
temperatures. Capillary condensation begins at lower densities in
the less dense aerogel, reflecting the smaller perturbation of the
helium vapor by the silica strands. The enhancement of the liquid
phase's density is smaller, but significant near the LVCP where it
is more compressible (Fig.~\ref{95to98Gel515K}). Measurements by
Tuliemeri \textit{et al}.~\cite{Tulimieri99} have also shown that
aerogel density affects the shape of isotherms far from the
critical point
--- they become sharper and flatter in high porosity aerogels.

\begin{figure*}
\includegraphics[width=\linewidth]{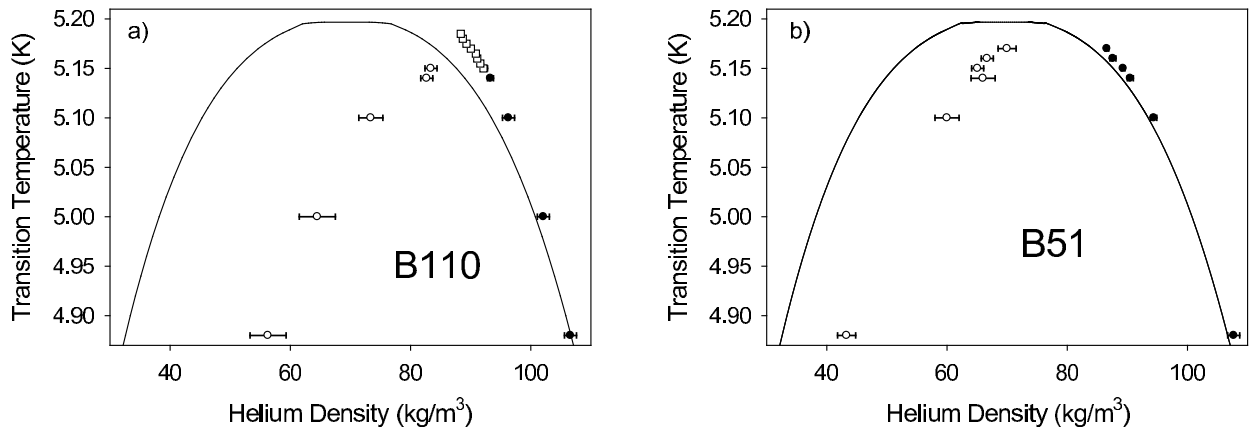}
\caption{\label{CCurve}Effective ``coexistence curves'' for helium
in aerogel B110 and B51 plotted using the closure points of
hysteresis loops (circles)  as markers for the high and low
density phases. In aerogel B110 additional data (squares) are
included which may indicate the completion of capillary
condensation at temperatures above the disappearance of
hysteresis.  Where error bars are not visible, they are smaller
than the data symbols.}
\end{figure*}

Most papers on the liquid-vapor transition in porous media include
plots of ``coexistence curves.''  Exactly what features to include
in such a plot, and even how to define those features, is somewhat
arbitrary; however, once the plot is constructed it does allow for
quick comparison to other experiments and to bulk fluid. To this
end, in  Fig.~\ref{CCurve} we plot ``coexistence curves'' for
helium in aerogels B110 and B51.  When hysteresis loops are
present, their closure points (which are always within the
accompanying uncertainty of any ``kinks'' in the adsorption
isotherm) are used to mark the onset and completion of
condensation in the aerogel. When there is no hysteresis, we
cannot identify a distinct low density feature, but over a limited
temperature range there is still a kink which may indicate the end
of capillary condensation and is a possible marker for the high
density ``liquid'' branch of a coexistence curve.
Figure~\ref{CCurve} compares these curves for our high and low
density aerogels, B110 and B51. The effective coexistence curves
in both aerogels are narrower than for bulk helium and are shifted
to higher densities. The changes are most significant for the
denser aerogel, B110, but the curve is not nearly as narrow as
Wong and Chan's coexistence curve for a similar porosity aerogel.
The narrowing of the B110 curve is, however, similar to that
previously seen for nitrogen\cite{Wong93} and neon\cite{tobyPhD}.

The behavior of fluids in pores near the liquid-vapor critical
point has often been discussed in terms of a ``capillary critical
point'' but there are a variety of definitions of this
point~\cite{Machin99,Thommes94,Morishige98}. It is sometimes
assumed to be the temperature where hysteresis disappears; in
other work it is thought of as a more fundamental thermodynamic
point which plays the same role as the liquid-vapor critical point
in bulk fluids. In the former case, properties like surface
tension are usually assumed to remain bulk-like.  In the latter
case, they may exhibit power law behavior close to the capillary
critical point.  In our measurements, the disappearance of
hysteresis does not correspond to a true critical point since, for
example, the fluid's compressibility does not diverge at this
temperature (the isotherm slopes do not approach zero).   Well
below the critical point, the isotherms become very flat, but the
hysteresis prevents us from measuring the equilibrium
compressibility to see if it diverges at some lower temperature.
It is clear that our isotherms and ``coexistence curves'' like
Fig.~\ref{CCurve} cannot be analyzed in terms of equilibrium
critical exponents, nor can we assign unique ``capillary critical
points'' for helium in our aerogels.

Although we cannot access the equilibrium states of the system
while it exhibits hysteresis, the simulations of helium
condensation in 95\% porosity
aerogels\cite{Detcheverry03,Detcheverry04} can probe the
energetics of states within the hysteresis loop.  The fluid
density in the equilibrium states showed a discontinuous jump in
fluid density at a single value of chemical potential as one would
expect with liquid-vapor coexistence. However, when the
simulations raise and lower the chemical potential to fill and
empty the aerogel (the analog of us varying the pressure in our
measurements) they also show metastable states and hysteresis. The
qualitative features of the simulated hysteresis loops closely
resemble our results and even the sizes of the loops are similar.
Furthermore, the simulated and experimental loops evolve in the
same way with temperature; roughly rectangular loops become
narrower and more triangular as the temperature is raised and
disappear slightly below the LVCP. At higher temperatures the
simulated isotherms have finite slopes, as do our experimental
ones.  If there is an equilibrium liquid-vapor transition, it must
be masked by the hysteresis loops at temperatures far from the
LVCP and disappear before the hysteresis loops do.

Although the simulated aerogels all have porosities of 95\% or
less, the porosity dependence of the isotherms also qualitatively
resembles the behavior in our experiments.  Loops become more
rectangular as the density of the gel decreases, and the
disappearance of hysteresis occurs closer to the bulk critical
temperature. Since the simulations which showed this behavior use
mean field calculations, it does not appear that critical thermal
fluctuations are needed to explain the evolution of the hysteresis
or its disappearance below the bulk critical temperature.  If true
thermal critical behavior is observable for fluids in aerogel,
then it must be very subtle.  This is also consistent with the
recent neutron scattering study of CO$_2$ in
aerogel\cite{Melnichenko04} which did not see any evidence of a
diverging coherence length near the LVCP.

Many of the features of our adsorption isotherms are similar to
those associated with capillary condensation in denser media,
despite the lack of well defined pores in aerogel.  In dense
media, fluid can be pictured as being adsorbed on the walls of
archetypal pores (cylindrical, slit, or ink bottle, for example).
As fluid is adsorbed, a liquid-vapor interface forms with a
negative curvature which shifts the equilibrium conditions and
causes liquid to condense below the bulk saturated vapor pressure.
This behavior is described quantitatively by the Kelvin equation,
which can be written as:
\begin{equation} P_0 -P_v = -C \gamma
\left( \frac{v_l}{v_v-v_l} \right)
\end{equation}
where $C$ is the curvature of the liquid-vapor interface, $\gamma$
is the surface tension of the fluid, $P_0$ is the saturated vapor
pressure at a given temperature, $P_v$ is the pressure at which
capillary condensation occurs, and $v_l$ and $v_v$ are the molar
volumes of the liquid and vapor phases respectively.  All of these
terms are temperature dependent except for the curvature of the
interface, which is assumed to remain well--defined and constant.
However, aerogels are best described as a tenuous network of
silica strands, not as a set of regular pores and it seems
unlikely that condensation would involve a liquid-vapor interface
with a constant negative curvature --- the mean field
simulations\cite{Detcheverry03,Detcheverry04} discussed above do
not show evidence for a constant curvature interface.

\begin{figure}
\includegraphics[width=\linewidth]{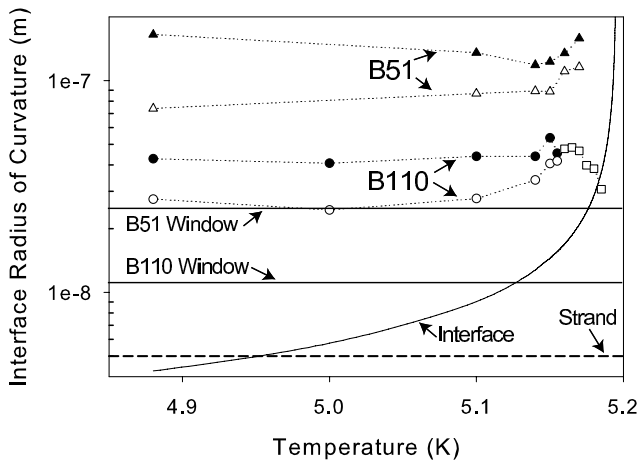}
\caption{\label{GelCurvaturePlot}Effective radii of curvature for
the helium liquid-vapor interfaces during adsorption (solid
symbols) and desorption (open symbols) in aerogels B110 (circles)
and B51 (triangles), from the Kelvin equation. The approximate
thickness of the aerogel strands and the bulk liquid-vapor
interface thickness are included on the plot, as are the window
sizes of cubic networks with densities equivalent to B110 and
B51.}
\end{figure}

The depression of the condensation pressure in our measurements
(filling at P/P$_0 < 1$ in Figs.~\ref{95Gel514RateHyst}
to~\ref{98GelHighT3Isotherms}) does scale roughly with the bulk
surface tension of helium, as one would expect for capillary
condensation.  If one assumes that condensation of helium in
aerogels can be well described by the Kelvin equation then we can
extract an effective curvature of the liquid-vapor interface which
is responsible for the lower condensation pressure.  To convert
curvature to a radius one must assume a shape for the liquid-vapor
interface, e.g. hemispherical. The effective radii for
hemispherical liquid-vapor interfaces in our aerogels (B110 and
B51) are shown in Fig.~\ref{GelCurvaturePlot}.
The radii were extracted from the filling and emptying pressures
using the Kelvin equation and assuming bulk values for surface
tension and molar volumes, Also included in the figure are several
other relevant length scales: the thickness of the aerogel
strands, the thickness of the bulk liquid-vapor interface, and
effective ``window sizes'' for models of aerogel as a cubic
lattice of intersecting cylinders, with the cylinder thickness and
sample density chosen to reflect our
samples~\cite{Scherer1998-399}.

The average low temperature (i.e. less than 5.150K) values for the
radius ``r'' extracted from the Kelvin equation are summarized in
Table~\ref{curvaturetable}. \begin{table}
\begin{center} \begin{tabular}{llc}
\textbf{Sample} & Branch & r\\ \hline \textbf{B110}  &  Filling  &
43nm \\ &Emptying& 28nm \\ \textbf{B51} & Filling & 135nm \\ &
Emptying & 85nm\\ \end{tabular}
\end{center} \caption{\label{curvaturetable} Effective radii of
curvature of the liquid-vapor interface extracted from the Kelvin
equation assuming a hemispherical meniscus.} \end{table} If the
interface has the same shape during filling and emptying, then its
effective radius of curvature must be roughly 50\% larger on
filling than emptying, but this difference more likely reflects
different interface shapes during adsorption and desorption.  For
desorption it may be possible to interpret the effective radius of
curvature as a ``breakthrough radius'' for the penetration of
vapor into the aerogel voids, but it is unclear what the effective
radius during adsorption could represent.

The thickness of a bulk liquid-vapor interface is given
by\cite{Bonn92} $ L = 3.64 \xi $ where $L$ is the interface
thickness and $\xi$ is the fluid correlation length calculated as
a function of reduced temperature, t, using $ \xi = \xi_0
t^{-0.63} $ where~\cite{Roe78-91} $\xi_0 (^4He) = (1.8 \pm 0.3)*
10^{-10}m$. When the interface thickness is less than the
thickness of the aerogel strands, it is reasonable to treat the
helium as a bulk fluid with a well defined meniscus. However, once
the interface exceeds the strand size, this picture breaks down.
The aerogel strands may not pin the interface and the effective
surface tension may differ from its bulk values. It is therefore
surprising that the filling and emptying of the aerogels scales so
well with bulk surface tension. Very close to the critical point,
where hysteresis disappears, the bulk helium interface thickness
is much larger than the strand diameter and is approaching the
size of the average void in the aerogel.  The effective interface
radius of curvature changes rapidly in this temperature regime and
it no longer makes sense to think of the system in terms of pores
with a curved liquid-vapor interface. It is probably better to
picture the aerogel as a random perturbation affecting the fluid
phases in some averaged way.

Although our adsorption isotherms share many qualitative features
with those in denser media, referring to the process as capillary
condensation is somewhat misleading since aerogels do not have
well defined pores.  Instead, one can picture the helium as
condensing around the strands and especially at their
intersections.  As adsorption proceeds, the fluid would fill
regions with the highest strand density first, as in the
simulations\cite{Detcheverry03,Detcheverry04}.

Since the aerogel silica backbone is so tenuous, it is easily
deformed by small forces, including the surface tension that
drives capillary condensation.  In fact, one can see the effect of
surface tension on the lower density aerogel in
Fig.~\ref{95to98Gel488K}.  The high pressure side of the B51
isotherm has a negative slope after the completion of capillary
condensation; this is due to surface tension compressing the
sample and changing the area and separation of the capacitor
plates.

We have studied the compression of aerogels by surface tension in
more detail using a different technique\footnote{T. Herman, J.
Day, and J. Beamish, to be published}.  At 4.880K, aerogel B51 is
compressed by up to 2\% by helium's surface tension during
desorption; B110, with its higher elastic constants, is compressed
by less than 0.2\% at the same temperature.  In that study we were
also able to measure the bulk moduli of our aerogels to be 0.4MPa
and 0.04MPa for B110 and B51 respectively.

\section{Summary}

We have investigated the adsorption and desorption of helium near
its LVCP in silica aerogel, a matrix with tunable density.  Our
measurements were made at temperatures far from the critical
point, where a well defined liquid-vapor interface exists, and at
temperatures close to the critical point where the fluid is
expected to sense the aerogel in an averaged way. Nowhere did we
see an unambiguous equilibrium first order transition, nor did we
see critical behavior dominated by thermal fluctuations. The very
flat low temperature hysteresis loops in aerogel may reflect an
underlying equilibrium phase transition in this system, but it is
not experimentally accessible because of metastable states during
filling and emptying.  Many features of our isotherms could be
described in terms of capillary condensation, although this
picture becomes less applicable as the liquid-vapor critical point
is approached. It is also unclear how the picture of capillary
condensation caused by a liquid-vapor interface with a
well--defined radius of curvature can be applied to aerogels,
which have structure over a wide range of length scales.  There is
no evidence for a constant radius of curvature in the adsorbed
fluid, and yet capillary condensation occurs over a very narrow
pressure range --- especially for very high porosity gels such as
B51.

%\appendix
%\section{}

\begin{acknowledgments}
We would like to thank Moses Chan, Etienne Wolf, Norbert Mulders
and Martin-Luc Rosinberg for helpful comments and discussions.
Funding for this project was provided by NSERC.
\end{acknowledgments}

%\bibliography{isothermsbib}

\end{document}